\documentclass{article}
\usepackage[T1]{fontenc}
\usepackage{lmodern}
\usepackage[ansinew]{inputenc}
\usepackage{graphicx}
\usepackage{epsfig}
\usepackage{bbm}

\begin{document}

\title{Nancy Cartwright and the logic of Quantum Mechanics }

\author{P. Lederer \\
Directeur de Recherche honoraire au CNRS\\
14 rue du Cardinal Lemoine, 75005-Paris, France\\
33(0)662984051}

\maketitle
\newpage
\begin{abstract}
This paper deals with Nancy Cartwright's views on the measurement problem in Quantum Mechanics, as exposed in her book {\it{How the Laws of Physics Lie}}. She does not accept the logic of Quantum Mechanics. It is argued  that her proposals, which are at variance with many facts results and epistemics of Quantum Mechanics are the result of her choice of classical logic, which leads her to propose  the transition rate as the fundamental object of Quantum Mechanics. I argue that this is incorrect. The positions which Nancy Cartwright defends on  the reduction of the wave packet do not address the fundamental issue, i.e. the duality of a world where quantum and classical objects coexist and interact. I suggest that the main problem with Nancy Cartwright's positions is her difficulty in accepting that the contradiction at the basis of Quantum Mechanics, i.e. the simultaneous corpuscular and wave-like nature of quantum objects, is a fact of nature. Recent experiments, described in this paper,  shed a new light on the foundations of Quantum Mechanics and on the topic of this paper. The limits of the no-contradiction principle are discussed; modern  dialectical materialism is argued to offer  a useful framework for the interplay between knowledge and reality.
\end{abstract}

\section{Introduction}
Classical physics is based on the notion of trajectory of a material object. A complete description of the latter is obtained when all coordinates and momenta of all its parts are simultaneously given. Non Linear Classical Mechanics has added to the classical notion of a trajectory that of the chaotic character of certain solutions of the equations of motion, whereby sensitivity to initial conditions turns trajectories in deterministic but unpredictable ones.

At the basis of the necessity to abandon classical physics is the observation that atoms are stable. In classical physics, electrons orbiting around a nucleus would radiate electromagnetic waves continuously, and collapse on the nucleus: atoms would not be stable.

Another basic experimental observation is that a sufficiently focused electron beam  with convenient energy, if  hitting a crystal, leads to an interference pattern on a screen, much like that observed for the diffraction figure of electromagnetic waves. Thus, under certain conditions, the behaviour of material particles exhibits  wave-like features. This phenomenon is radically at variance with the classical notion that a corpuscular (discontinuous) entity cannot have a wave like (continuous) behaviour. 

As emphasized in many Quantum Mechanics (hereafter QM) text books, and in particular in the first chapter of {\it{ Quantum Mechanics}} by Landau and Lifchitz \cite{landau}, QM has thus  to be based on motion concepts fundamentally different from those of classical mechanics: ``{\it{In QM, the notion of a trajectory does not exist. This is expressed by the uncertainty principle, which Heisenberg discovered in 1927 \footnote{ Like\normalfont many physicists nowadays, I would prefer ``undetermination `` to ``uncertainty''  }}}''. This book first appeared in 1966, but it is not discussed in Nancy Cartwright's book \cite{cartwright} (published in 1983) {\it{How the Laws of Physics Lie }}. In particular  the last chapter entitled : {\it{How the Measurement Problem is an Artefact of Mathematics }}is of interest in this context.

In this paper, I will examine the various theses developed in that chapter of her book. I will argue that her position that {\it{``transition probabilities are fundamental''}} although  correct in certain areas of physics, ignores a vast amount of facts which prove that this position is incorrect in general. As discussed  below, I claim that this is due to  the failure to accept the {\it{ funny logic}} of QM; furthermore, I will argue that Cartwright does not identify correctly one of the last fundamental unsolved  problems of QM nowadays: the duality of a world where quantum behaviour coexists with classical behaviour. A problem on which results obtained in the last few years \cite{haroche,wineland} have allowed perhaps decisive progress. This should   be taken into account in the philosophical debate about QM.

The question of how classical behaviour emerges from a set of interacting quantum particles is an actively debated one \cite{laloe}. I am not interested  in the relevant  theoretical issue here. This paper is not discussing decoherence.  The latter  may suppress interferences, but is well known  not to solve the question of the ``single pointer position'' of the classical object, as described below. The aspect I am underlining here is that the recent experiments mentioned in the last paragraph provide new evidence for the classical /quantum duality of our world, as well as for the Copenhaguen School interpretation of QM.

 The structure of the paper follows at first that of Nancy Cartwright's chapter on the measurement problem  in her book (reference \cite{cartwright}). Section \ref{measure} states the position of the measurement problem in QM, based on exact quotations from active members of the Copenhagen School,  and lists Cartwright's theses. Section \ref{defence} discusses the topic of the quantum state. Section \ref{transitions} deals with Cartwright's main thesis about transition probabilities in QM. Section \ref{artefact} discusses the notion of the measurement problem as an artefact of mathematics, as proposed by Cartwright. Section \ref{duality} deals with the duality of the quantum and of the classical world.

One crucial issue discussed in the course of this study is that of the coexistence of contraries within objects of nature. I have felt useful and necessary to add to the study of Nancy Cartwright's proposal a discussion (section \ref{contradiction})  on the scope,  validity and  limits of the aristotelian principle of non contradiction, and on some basic notions of dialectical materialism.

\section{The measurement problem} \label{measure}

The  title of this chapter, i.e. {\it{How the measurement problem is an artefact of the mathematics}},  sets the stage. Referring to von Neumann's work of 1932 \cite{neumann}, there are two kinds of evolution in the quantum theory; one is governed by Schroedinger's equation, which is continuous and deterministic; the other is the {\it{reduction of the wave packet}}, which is discontinuous and indeterministic.  Cartwright's goal along all this chapter is to deny this duality. 
The following is a quote of the introduction to the chapter\footnote{P. 163 of reference \cite{cartwright}.}:
\begin{quote}{\it{ ...Most of the time systems are governed by the Schroedinger equation. Reduction occurs when and only when a measurements is made. How then do measurements differ from other interactions between systems? It turns out to be very difficult to find any difference that singles out measurements uniquely. Von Neumann postulated that measurements are special because they involve a conscious observer.... I think that this is  the only solution that succeeds in picking measurements from all other interactions. It is clearly not a satisfactory solution...\footnote{This passage shows that the author adopts a rational position and rejects the idealistic interpretations of QM. I fully agree with her on this, which will not be addressed further in this paper.} I will argue here that the measurement problem is not a real problem. There is nothing special about measurement. Reductions of the wave packet occur in a variety of other circumstances as well, indeterministically and on their own, without the need for any conscious intervention...On the conventional interpretation, which takes position probabilities as primary, \textsf{\textbf{ quantum propositions have a peculiar logic}}\footnote{bold type  here is by me, as in the other bold type quotations in this paper.}, or a peculiar probability structure or both. \textsf{\textbf{But transition probabilities where reduction of the wave packet occur, have both a standard logic and a standard probability. They provide a non problematic interpretation of the theory.}} 

The proposal to develop an interpretation for quantum mechanics based on transition probabilities seems to me exactly right.}} \end{quote}

In her attempt to justify this interpretation, the author encounters however  a difficulty: \begin{quote}{\it{Two kinds of evolution are postulated. Reductions of the wave packet are no longer confined to measurements, but when do they occur? If there are two different kinds of change, there must be some feature which dictates which situation will be governed by Schroedinger's law, and which by the projection postulate.  }}\end{quote} Eventually, she urges that \begin{quote} {\it{the two evolutions are not different in nature; \textsf{\textbf{their difference is an artefact of the conventional notation. }}}}\end{quote}

In the following, I will describe a different proposal, which was actually clearly exposed by brilliant members of  the Bohr Copenhagen school, such as Lev Landau. Some readers will frown and accuse me of being conservative and conformist. It turns out that very recent experiments, which will be discussed in section \ref{duality} add new evidence in favour of the Copenhagen interpretation, as detailed below. Disregarding those new experiments would be a  mistake. The proposal formulated in this paper is based on the full recognition that QM is a bona fide example of a different logic, a logic which is at work in nature, and which is at  variance with the traditional Aristotle prohibition of contradictions; a logic which led Bohr to adopt as motto: \textit{ Contraria sunt complementa.}

 My view is that Nancy Cartwright's failed attempt at denying  the duality at the core of QM is based in part on the failure to recognize the nature of this duality, and in part on the influence on her thinking of the  dominant rejection of a logic which admits contradictions in epistemics, as imposed by contradictions within objects of nature\footnote{In that respect, Nancy Cartwright is part of a large family of positivist  or neo-positivist philosophers, such as Duhem \cite{duhem}, whom she quotes approvingly in her book, Carnap \cite{carnap}, Kuhn \cite{kuhn}, Van Fraassen \cite{fraassen}, Popper \cite{popper}, Lakatos \cite{lakatos}, etc.. Those will not be discussed further in this paper}.

\section{In defence of reduction of the wave packet}\label{defence}

In her section ``{\it{In defence of reduction of the wave packet }} ``, the author first describes in more details the difficulty with the measurement problem. This difficulty is met by all authors who believe that the Schroedinger equation is the only fundamental equation, governing all objects of the universe, which usually leads to ideas of parallel universes, and in the end, to the position that there is no possibility of knowing the world \cite{espagnat}. Nancy Cartwright does not share this  position, but her proposal to solve the problem is unacceptable, as I will argue in this paper.

In this section, she describes the current reasoning about the interaction of a quantum particle -- such as an electron -- with a ``{\it{macroscopic}}'' body. Quoting Cartwright:{\it{\begin{quote}...It is possible for macroscopic objects to be in states with well defined values for all macroscopic observables... But interactions with microscopic objects bring them into superpositions. The electron starts out in a superposition with the apparatus in its ground state\footnote{In fact in an initial state, not necessarily the groud state.} Together the composite of the two finishes after the measurements in a superposition, where the pointer of the apparatus has no well defined position but is distributed across the dial...After the measurement has ceased, a new kind of change occurs. The superposed state of the apparatus-plus-object reduces to one of the components of the superposition. This kind of change is called the ``reduction of the wave packet'', and the principle that governs it is the ``projection postulate''.\end{quote}}}

For clarity, let me quote in the following how Landau and Lifchitz describe this \footnote{p. 32 of reference \cite{landau}.}, with an added crucial statement (hereafter in bold letters):{\it{
\begin{quote} Consider a system with two parts:  a classical apparatus and an electron (a quantum object). The measurement process is defined by their interaction, the apparatus changes from its initial state to another state; this change allows to study the electron state. The states of the apparatus are characterized by certain physical values, denoted by $g$, with eigenvalues $g_n$. For simplicity the spectrum of the $g_n$ is supposed to be discrete. The states of the apparatus are noted $\Phi_n(\xi)$, where $n$ corresponds to the apparatus ``pointer position'' $g_n$, and $\xi$ is the set of its coordinates; \textsf{\textbf{The classical character of the apparatus is expressed by the fact that we are certain that, at all time, it is in one of its known states $\Phi(n)$ with a certain determined value $g$}}.

Let $\Phi_0(\xi)$ the apparatus initial state wave function (before measurement) and $\Psi(q)$ a certain electron initial wave function ($q$ is the set of its coordinates). These functions describe independently the states of the apparatus and of the electron, and thus the initial wave function of the whole system (before interaction) is the product
\begin{equation}
\Psi(q)\Phi_0(\xi)
\end{equation}
When the apparatus and the electron interact, QM allows in principle to follow the time variation of the whole system wave function. After the measurement process, expanding the latter in terms of the apparatus complete set of wave functions, we get a superposition of the form:
\begin{equation}\label{superposition}
\sum_n A_n(q)\Phi_n(\xi),
\end{equation}
the $A_n(q)$ being certain functions of $q$ \footnote{See the introduction to reference \cite{landau} }

\textsf{\textbf{This is the time for the apparatus ``classicism'' to enter the stage, as well as the classical mechanics duality, as limiting case, and, at the same time, foundation of Quantum Mechanics}}. Thanks to the classical character of the apparatus the quantity $g$  has a determined value (``the apparatus pointer position'') at each time. Thus we can state that the apparatus$+$electron system state will not be described by a sum such as equation (\ref{superposition}), but by a single term which corresponds to the ``apparatus pointer position'' $g_n $, i.e.
\begin{equation}\label{projection}
A_n(q)\Phi_n(\xi)
\end{equation}
\end{quote} }}

The text then proceeds to demonstrating that the $A_n(q)$ have to be of the form
\begin{equation}
A_n(q) = a_n \phi_n(q)
\end{equation} 
the $\phi_n$  are the normalized electron wave functions, after the measure, and $a_n$ are the only constants which depend  on the initial state $\Psi(q)$\footnote{In this paper, I refer indifferently to the ``quantum state'' and to the ``wave function''. Technically, the wave function is the expression in real space, $<\vec{r}|\psi>$ of the quantum state  $|\psi>$ in Dirac notation. This has no bearing on the topic of this work.}
\begin{equation}
a_n= \int \Psi(q)\Psi_n^*(q) dq
\end{equation}
where the $\Psi_n(q)$ form a complete set of orthonormal functions  which depend on the measurement process.

The bold type sentences seem to me the fundamental issue which allows to understand the dual combination of the Schroedinger equation and of the projection postulate about the measurement process. Namely the \underline{postulate} that there exists in the world objects which obey classical laws of motion, and others which obey QM. This postulate is coherent with the fact that {\it{classical mechanics is \underline{both} a limiting case, and, at the same time, the  foundation of Quantum Mechanics.}} This issue is addressed later on in this work. I argue that the Copenhagen school has correctly suggested  the solution, i.e. quantum versus classical,  which, incidentally, is not synonymous with the opposition between microscopic and macroscopic, as emphasized in ref. \cite{landau}. Recent experiments \cite{haroche, wineland} have shed new light on this issue, and led to the Nobel Prize  awarded to Serge Haroche and David Wineland. 

In the remaining parts of the  section 1 of the chapter, Cartwright attempts to show that the Schroedinger equation cannot be the basic equation of the world, since interactions lead to vastly complicated entangled states. Her conclusion of this section is that reduction of the wave packet must occur quite generally. But she does not acknowledge that ``measurement'', and ``reduction of the wave packet'' may occur anytime a quantum  entity interacts with a \underline{classical} one. She rather ascribes this to interaction with \underline{macroscopic} objects. This is not a minor point. Macroscopicity is not the issue. There are macroscopic quantum objects, such as superconducting states, Quantum Hall states or neutron stars, etc.. The issue is quantum vs classical. In the Haroche group experiments, the classical object is reduced to a hundred atoms!

\section{Are transition probabilities  fundamental?}\label{transitions}

The section 2 of the chapter is entitled ``{\it{Why transition probabilities are fundamental}}''.
It starts by an account of the two-slit experiment: electrons are directed from a source  to a photographic plate but there is an opaque screen between the source and the plate. The screen is pierced by two conveniently spaced slits, noted 1 and 2. The amplitude  to move from the source to a given point $x$ on the screen is $a_1(x)$ if only slit 1 is open. It is $a_2(x)$ if only slit 2 is open. When both slits are open, the total amplitude is $S(x)=a_1(x) + a_2(x)$. Both amplitudes are derived from the Schroedinger equation. the sum $S$ is the standard QM result due to the linearity of the Schroedinger equation. The probability for an electron to hit the screen at $x$ when both slits are open is $|S|^2 = |a_1(x)|^2 + |a_2(x)|^2 + a_1(x)^* a_2(x) + a_1(x) a_2(x)^*$. The last two terms are interference terms: when both slits are open, the result for $|S|^2$ is not the sum for the processes when only one slit in turn is open. This example is a simplified version of the electron diffraction by a crystal: it accounts satisfactorily for the experimental results; it shows that consideration of probabilities only, inspired by the conventional classical logic, is an error. The possibility of interferences is a fact of life as far as quantum entities are concerned. Cartwright, on the other hand, is not satisfied, because her logic prohibits the possibility that a single particle traverses the screen simultaneously through both slits. It seems to me that  this reflects her difficulty in accepting that the  classical trajectory  notion breaks down for quantum entities. She spends several pages reviewing various proposals by philosophers, who, like her, do not want to consider amplitudes or wave functions as a fundamental object of QM, and who struggle  with the interference term which seems to reflect a ``{\it{funny logic}}'', at variance with the traditional one based on classical mechanics trajectories. In the course of this discussion she reveals a misconception about  QM rules for addition of  amplitudes. She writes: {\it{ \begin{quote} ...It supposes that the electron passes through neither one slit nor the other when we are not looking. When we do look, suddenly, there, it is either at the top slit or at the bottom. What is special about looking that causes objects to be at places where they would not have been otherwise?   \end{quote}}}

 ``Looking through which slit the electron passes'' is detecting it through some interaction with some material means. For example, a photon scattered by the electron at one slit. In that case, as explained in the first pages of Feynman's  Lecture Notes on QM \cite{feynman}, one cannot sum the amplitudes for passing through slit 1 and slit 2, because the final states at the screen  are not identical: there is a scattered photon in one final state, and none in  the other. In fact, whenever an inelastic process  occurs along one  path, the corresponding amplitude cannot be added to the other amplitude, and no interference effect results from this path. Summing over amplitudes for a given process requires the initial state and the final state for all paths to be identical. When writing a paper about the measurement problem, it would seem necessary to acknowledge this fact...

However, unhappy about all proposals to understand the two-slit experiment and the {\it{funny quantum  logic}} exhibited by the interference term, the author proposes\footnote{p. 179, reference\cite{cartwright}} {\it{\begin{quote} a more radical alternative. I want to eliminate position probabilities altogether, and along with them the probabilities for all classical dynamical quantities. The only real probabilities in QM, I maintain, are transition probabilities ...

I shall illustrate this with a couple of examples: the first, exponential decay; and the second, the scattering of a moving particle by a stationary target; etc.. \end{quote} }}

Let us admit that  exponential decay, or particle scattering by a stationary target, or maybe a number of other phenomena can be dealt with by looking at transition probabilities.  Using a few examples as proof of  a universal statement ( {\it{The only real probabilities in QM are transition probabilities }}) is not acceptable. Should not Hume's warnings \cite{hume} about induction prevent us from such a  reasoning? In the following I'll give counter examples which show that no generalization of the sort Cartwright advocates is allowed.

She adds furthermore\footnote{p.186, ibid.}: {\it{\begin{quote} I have been urging that the interpretation of quantum mechanics should be entirely in term of transition probabilities.  When no transitions occur, $\psi$ must remain uninterpreted or have only a subjunctive\footnote{misprint? subjective seems more correct.} interpretation.   \end{quote} }}
This is another incorrect stand, which I will show in a later section to be contrary to a number of examples.
Another disputable position is quoted,  in line with similar statements in the chapter \footnote{p. 182, ibid.}:
``Henry Margenau has long urged that all quantum measurements are ultimately measurements of position''. The only canonical conjugates Cartwright ever considers in this chapter are momentum and position, which are related through the indetermination relation $[q,p]=i\hbar$. What about the transverse components of the angular momentum, related by $[l_x,l_y]= il_z$? What about the relation between the guiding center coordinates $R_x, R_y$ of the electron motion under magnetic field in two dimensions \cite{lederer}, which are  related by $[R_x,R_y]= il_B^2$, where $l_B$ is the magnetic length? What about the undetermination relation connexion of  energy and time\footnote{Energy and time are not canonical conjugate. I thank Jean-Noël Fuchs for a remark about this.}? 
What about the canonical conjugation of phase $\phi$ and number $N$ which is at the basis of phase coherence in coherent optics, superconductivity \cite{BCS}, etc..? Whenever two dynamic quantities are expressed as $Q$, and $\alpha d/dQ$ where $\alpha$ is a constant, such quantities are canonical conjugates. In other words, eliminating the ``funny logic'' of QM by eliminating position probabilities leaves unsolved the ``funny logic'' at work in all canonically conjugate quantities.

\section{Is the measurement problem an artefact of mathematics?} \label{artefact}

The last section of the chapter is entitled : {\it{ How the measurement problem is an artefact of the mathematics}}.
It starts  with the following statement;\begin{quote} {\it{Reduction of the wave packet goes on all the time, in a wide variety of circumstances. There is nothing peculiar about measurement...The measurement problem has disappeared. But it seems that another has replaced it. Two kinds of evolution are still postulated: Schroedinger evolution and reduction of the wave packet. The latter is not restricted to measurement type situations, but when does it occur? What features determine when a system will evolve in accord with the Schroedinger equation, and when its wave packet will be reduced? I call this\footnote{P. 196, ibid. Underlined by me.} \underline{the characterization problem.}}} \end{quote}

Contrast this with the very first pages of reference \cite{landau}: \begin{quote} {\it{ A measurement, in QM, describes any interaction process between  a classical entity and a quantum one.

We have defined the apparatus as a physical entity which obeys classical mechanics with sufficient accuracy. Such as, for example, a body with a sufficiently large mass. But it would be incorrect to deduce from this that macroscopicity is a compulsory feature of the apparatus...The role of apparatus may be played by a microscopic entity, since ``sufficient accuracy'' depends on the actual problem at hand. }} \end{quote}

Comparing the two quotations above, one concludes that both agree with the notion that reduction of the wave packet occurs all the time. However Nancy Cartwright  does not consider the quantum-classical duality as a relevant notion, contrary to the Copenhagen school, as expressed by the quotation in section \ref{defence}.

The section then proceeds by examining various attempts, such as dealing with the system sizes, to solve the ``characterization problem'', only to find out they do not succeed. Eventually, the author states: {\it{ \begin{quote} Sheer size cannot solve the characterization problem as I have laid out... it is a pseudo problem. The characterization problem is an artefact of mathematics. 
There is no real problem because\footnote{P. 198, ibid. Underlined by me.} \underline{there are not two different kinds} of evolution in QM. There are  evolutions that are correctly described by the Schroedinger equation, and there are evolutions that are correctly described by something like von Neumann's projection postulate. But these are \underline{not different kinds} in any physically relevant sense. We think they are because of the way we write the theory. \end{quote}}}

In the remainder of this chapter, the author attempts to justify this position by invoking quantum field theory and by suggesting that using non unitary evolution operators in the former would eventually prove her  point.

There is little doubt that quantum field theory is a useful tool to treat the physics of large numbers of identical particles such as fermions or bosons; statistical averages over a large number of particles governed by the Schroedinger evolution operator do not depend on the reduction of single particle wave packets. However, quantum field theory is of little help in the  physics of a small number of quantum particles. Nowadays, technological and fundamental science progress allows to measure  single atoms irradiated with a few photons  \cite{wineland} or to explore states with a few photons using single atoms \footnote{typically the number of photons is of order 10.} \cite{haroche}. It may be that the program described by Nancy Cartwright succeeds, but its results do not solve either the ``characterization problem'' or its origin, namely the duality that she claims she has eliminated, because she has not clearly identified its essence.

\section{The origin of duality: quantum world vs classical world} \label{duality}

The Copenhagen group has clearly expressed -- as shown by the quotation above from reference \cite{landau} -- the view that quantum mechanics cannot exist without the classical world. The introduction of this book \cite{landau} states {\it{\begin{quote} Ordinarily, a more general theory can be formulated in a logically closed way, independently of a less general one. Relativity theory, for example, is built on fundamental principles without resorting to newtonian concepts. As for the formulation of the fundamental principles of quantum mechanics, it is basically impossible without the intervention of classical mechanics...It is clear that in a system built  exclusively on quantum entities, there would be no possibility of formulating logically closed mechanics.\end{quote}}}

This statement deserves thorough consideration. Classical mechanics has been built over the years since Galilee. It embodies a large amount of theoretical and experimental successes, of successful predictions,  as well as infinitely many applications. Hamiltonians and Lagrangians are fundamental constructs of  classical physics; they are fundamental to develop any QM investigation. The theorist starts with the the expression of the classical Hamiltonian, or Lagrangian, of a system and proceeds by quantizing it, i. e. by introducing the conjugation relations between position and momentum, and/or between components of the angular momentum, etc.. 

However, do we fully  understand today  under what condition an entity will behave classically, so that the superposition due to its  interaction with a quantum one will end up in a single state, with the ``single indication of the apparatus''? The usual answer is that when the action of a system  is very large compared to $\hbar$, it behaves classically \cite{feynman2}. But this is not the whole story.  We know various examples of macroscopic quantum states, such as the BCS ground state \cite{BCS}, the quantum Hall states \cite{lederer, laughlin}, various superfluid states, Bose condensates, neutron stars, etc., so that macroscopicity cannot be the  universal answer. 

Recent experiments conducted in \'Ecole Normale Sup\'erieure in Paris \cite{haroche} shed a bright light on the issues discussed by Nancy Cartwright, and do not support her view that ``the laws of physics lie'', or, for that matter, that the ``measurement problem is an artefact of mathematics''. 
In those experiments,  a few photons are trapped in a cavity in  a coherent superposition state with a spread in the number $N$ of photons, and a fixed phase $\phi$, following the canonical uncertainty relation between phase $\phi$ and  number $N$, i.e. $\Delta \phi . \Delta N \approx 1$.
Conveniently prepared Rydberg atoms are then shot one by one  through the cavity where they interact with the electromagnetic field, which alters their energy levels depending on the number of photons which have interacted with the atom inside the cavity\footnote{See reference \cite{laloe} for more details}. This perturbation is analysed when the atoms emerge from the cavity, so that the number of photons in the cavity is determined. What is found is that this number first fluctuates with the first atoms; the distribution of states according to the number of photons  is found to  obey a Poisson distribution of probabilities; then the number of photons fluctuates less and less as more  atoms are shot through the cavity, until, after about 100 atoms, a single number is found to be stabilized. In other words, the quantum state of photons inside the cavity is projected by its repeated interaction with the atoms in a pure state with a fixed number of photons, while its phase becomes undetermined. The interpretation of this experiment is that the ``classical entity'' which projects the superposition of states into a pure state is the set of about 100 atoms which have been shot through the cavity. On top of this, the experiment allows to follow how this ``classicism'' gradually appears as the number of atoms increases.  Once a pure state is created, any later measurement finds it remains unchanged, as QM text books teach. Other experiments on entangled states made of two level atoms and superposed photon states allow to give flesh to a ``photon-made'' Schroedinger cat\footnote{An illuminating talk in French by Serge Haroche in Ecole Polytechnique in 2014 is highly recommended.}, etc.. 

Such results on single quantum entities built out of a few units of quantum particles vindicate the ``funny quantum logic'' which Nancy Cartwright urges to forget in favour of the Boolean familiar classical logic of transition probabilities. Bohr's motto is an expression which states that  contraries  ``complement'' each other; I interpret it\footnote{Such was not Bohr's position.}  as  a statement about  the contradictory unit formed by particle and wave, i.e. a contradiction between continuity and discontinuity. Bohr's motto is quite different from Aristotle's statement : ``contraries exclude one another''.
In QM, both terms coexist, and either term of the contradiction dominates depending on the experimental conditions. Wave character and corpuscular character  of a quantum particle are two contradictory aspects which are simultaneously present and unavoidable in the rational analysis of QM experiments. In fact, the notion of wave packet -- a superposition of plane waves which may result in a localization of the particle -- shows that either aspect may dominate depending of the superposition: an infinite sum of plane waves describes a point-like particle; the two contradictory features become, in  some sense identical. In fact, in my view, the main reason why Nancy Cartwright is proved wrong so convincingly by experiments is her philosophical blindness to the ``funny  logic'' which admits inner contradictions within  entities as a possible (general?) mode of existence of reality\footnote{See for example references \cite{seve} for a  more detailed discussion on the dialectics of nature}. The quotation in bold type in section \ref{measure} exhibits a solid faith in ``standard logic'', as a decisive cause for her ``defence of transition probabilities''. Standard logic is the Aristotelian logic which prohibits contradiction within the thing: ``A cannot be non A''. The value and limits of validity of this ``no-contradiction'' principle is discussed in more details in section \ref{contradiction}.

The  point of view following which the only equation governing the world is the Schroedinger equation (see for example references \cite{espagnat,everett}) is somewhat different from Nancy Cartwright's. In this ``Schroedinger only'' point of view, infinitely many  worlds coexist and develop in infinitely many superposition states. This position also denies the duality discussed above. It disregards the impossibility of funding QM without classical mechanics. It results in the impossibility of establishing any truth about the world we live in. Such theories and philosophies will have a hard time arguing away the results described above. The existence of objects obeying classical mechanics is a theoretical and experimental achievement of mankind which should not be dismissed because of the existence of objects obeying  QM.

Another debatable issue is Nancy Cartwright's stand  that when there are no transition probabilities, the wave function cannot be given any interpretation. To begin with, the existence of quantum states is dictated by the breakdown of the classical notion of trajectory at the microscopic level. The quantum state, which contains necessarily less information than the simultaneous existence of position $q$ and momentum $p$ in classical mechanics, replaces those two entities as the best possible theoretical entity when the former do not exist simultaneously. No transition probability can be inferred without the notion of quantum state. But is the quantum state, and its space description, the  wave function,  only a ``theoretical entity''? I urge to consider the wave function as a bona fide real entity. Consider the structure of molecules, which can now be observed in detail one by one by various techniques. Not only are their shape and spatial structure dictated
by the shape of the atomic wave functions, but knowledge of the latter allows to build taylor made atomic complexes with the desired shape  or chemical activity. Hacking \cite{hacking} \footnote{more than hundred years after Marx...}  considers practice as a  decisive test of reality for theoretical entities. If he is right -- as I believe he is -- atomic wave functions have to be considered as facts of life, which are used practically to create specific molecules.\footnote{The question of many body states which are mathematically written in very large space dimensions requires a special discussion, which is not treated in this paper.}

 Macroscopic quantum objects, such as a piece of superconducting metal, are endowed with a macroscopic coherent ground state wave function which has a fluctuating total number of particles and a fixed phase\footnote{The superconducting phase breaks electromagnetic gauge invariance.}. The latter  leads to such devices as SQUIDS \cite{tinkham} which are spectacularly accurate, among other things, in measuring very small magnetic field intensities\footnote{such as magnetic fields produced by the nervous system in the human brain.}, in accounting for the Meissner effect, or the occurrence of quantized filaments of flux (vortices) in a superconductor submitted to a magnetic field, etc.. All such effects empirically well known and theoretically well understood do not depend on transition rates, but on the structure of the macroscopic wave function. The basic explanation of the  Quantum Hall Effects is based on the wave function  derived by Laughlin \cite{laughlin}. The latter, in the case of the fractional version of the Hall effects, led to the prediction (and subsequent observation) of fractionally charged particles in the fractional quantum Hall state, by straightforward examination of the wave function phase. Another example is the vanishing longitudinal resistivity  which coincides with the quantum Hall transverse resistivity plateaux: it is due to the absence of any scattering process (i.e. any transition) for the electron in a chiral surface state...Those examples are sufficient to establish that Nancy Cartwright's position on the lack of meaning of the wave function when no transition rates are available is not supported by a number of successful explanatory theories with practical applications. 

\section{Limits of the no-contradiction principle} \label{contradiction}

At this stage, I feel it is necessary to spend a few lines on the question of the ``contradiction within objects of nature'' which I have mentioned too briefly above. If what I am writing below  appears irrational, and in particular if the notion of unity of contraries in the epistemics  and ontology of  objects of nature still appears as unacceptable, my analysis of the reasons for Nancy Cartwright's positions on QM will fail to convince too. There is currently a large sector among philosophers and in particular philosophers  of science who consider the ``no-contradiction principle'' established by Aristotle has such a validity that any mention of coexistence and - worse -, identity  of contraries in objects of nature is immediately rejected by many, with no hesitation or discussion. There are very good reasons for this position. But there are also good reasons to reconsider it in view of some arguments I discuss now.

The history of  dialectical thinking is about three thousand years old \cite{seve}.That some truth may be expressed in a contradiction is at the heart of the most ancient philosophies in China, India, Greece, from Lao-Tseu to Heraklite \footnote{Example from Heraklite: `` It is impossible to bathe twice in the same river''}. This was radically dismissed by Aristotle \footnote{Aristotle : ``the same cannot belong and simultaneously not belong to the same simultaneously and under the same connection''(ontology); ``contradictory statements cannot be simultaneously true'' (logic); ``nobody can believe that the same could be simultaneously be and not be''(psychology)}, in the name of a logic he established \cite{aristo}. Dialectical thinking  was banished in the western world   as detrimental to social order, or even the possibility of communication. A crucial point is that the no-contradiction principle is based on an ontological postulate, i.e. the invariance of the essence. Masked by the universal sensible change  in our sublunar world, the latter is  nevertheless -- says Aristotle --   the ultimate truth of the being\cite{seve,aristo}. 

Then Kant, at the time of the French Revolution \cite{kant} finds out that the century old efforts of metaphysics result in fundamental and unsolvable contradictions, which he dubbed antinomies. The latter, says Kant, are a sign of a fundamental limitation of our understanding. 

Then Hegel \cite{hegel} takes a bold new stand: if all quest for truth inevitably results in a contradiction, then contradiction is the truth!

For Hegel, dialectics is not illogical, it is logic developed beyond the limits of aristotelian logic. The no-contradiction principle appears to be relevant for the invariant, the  inert thing, but it is in great difficulty to think connections and processes, as illustrated by the problems posed since ancient times by the simple motion of an arrow. Hegel's ideas caused enormous interest in Europe in the 19th century, as well as fierce opposition, in particular by the catholic church, for which Hegel is a pantheist, or worse, an atheist\footnote{I cannot review here those developments, for want of space. Interested readers are invited to read, for example, the recent book by S\'eve (chapter IV) \cite{seve}.}. The third Empire in France banishes Hegel's ideas from Academia. In turn the working class movement and the rise of socialist thinking  triggered  a revival of hegelian studies. Marx \cite{marx}  adopted hegelian dialectics, but rejected the idealist hegelian position following which the Concept  precedes reality; for him, dialectics appear in the theory of knowledge as a result of dialectics in the objective world. Engels \cite{engels1} develops dialectical materialism, formulates laws thereof; at first dialectics is for him objective inasmuch as it is imposed by reality to our subjective logic; later  he reaches a disputable position: dialectical materialism becomes, in a  pure ontological way  {\it{the science of the general laws of motion of the external world as well as those of human thought}}\cite{engels2}: this thesis dismisses the essential epistemic aspect of dialectics and eventually opens the way to the catastrophic version of stalinist dogmatism.

After the defeat of nazism, dialectical materialism became an important philosophical current. Most serious philosophers in Europe, Asia and elsewhere adopted it, in one version or another. This success led to its demise. It suffered a severe blow when it was used as official state philosophy in the USSR.  Much to  the contrary, nothing,  in the founding philosophical writings \cite{marx,engels1,lenin} allowed to justify turning them into an official State philosophy.  This produced  such catastrophies as  the State support for Lyssenko's theories, based on the notion that genetics was a bourgeois science, while lamarckian concepts were defined at the government level as correct from the point of view of a caricature of dialectical materialism. It is understandable that such nonsense in the name of a philosophical thesis turned the latter into a questionable construction in the eyes of many.

Dialectic materialism itself is an open system, which has no lesson to teach beforehand about specific objects of knowledge, and insists \cite{engels1} on taking into account all lessons taught by the advancement of science.

 It is perhaps time for a serious critical assessment of this philosophical thesis. The possibility of general theoretical statements  about the empirical world is not a negligible question. 

 Materialism gives a clear answer to the "`{\it{gnoseological problem of the relationship between thought and existence, between sense-data and the world...Matter is that which, acting on our senses, produces sensations.}}'' This was written in 1908 by Lenin \cite{lenin}. It may look too simple when  technology (such as that used in QM experiments) is intercalated between matter (the  electrons in the two slit experiment) and the screens on which we read their impact.  Technology or not, matter is the external source of our sensations. So much for materialism. Dialectical materialism adds  a fundamental aspect  i.e. that theories are led to analyze reality   in terms of  contraries which  coexist and compete with each other within things in Nature. Depending on which dominates the competition (contradiction) under what conditions, the causal chains originating from the thing and causing phenomena will take different forms, which are reflected in theories. Epistemics and ontology are intimately intertwined \cite{seve2}. 

Consider an example of how formal logic and dialectical logic  complete and enrich each other: that of cause and effect. A moving billiard ball 1 hits a motionless billiard ball 2 which is thus set in motion. The motion of 1 is the cause for the motion of 2, which is the effect. Cause and effect are two contraries of a logic of identity: their meaning is clear, the relation is uni-directional: there seems to be no room left for contradiction. However, if the collision has caused the motion of 2, due to that of 1, the trajectory, energy  and  velocity of 1 have also been changed; to the initial causal relationship wherein 1 is a cause for 2 is added necessarily an inverse causal effect wherein 2 becomes  a cause for the motion of 1. The uni-directional causal relation we had first is turned immediately into a reciprocal relation: a cause leading to a consequence is in turn affected by the consequence turned into a cause itself. Can't we see here an example of unity of contraries? The classical logician will deny it, observing that there is no contradiction, but interaction: coincidence of two causal actions which remain distinct causal ones. However, how can we distinguish within the collision the causal action of 1 on 2 and that of 2 on 1?  

Another example is given by Aristotle's fundamental categories such as quality and quantity. In the classical logic, those two categories are clearly distinct and form a couple of well identified contraries. But there is  a host of empirical evidence that there is no such  dichotomy : quantity may transform into quality almost universally; think of all examples of spontaneous symmetry breaking, for example. More trivially, one sleeping pill  puts one to sleep, hundred  kill her, etc..

What I have discussed in this paper, about Nancy Cartwright's views on QM has its origin in the long lasting historical conflicting views since Newton and Fresnel between the corpuscular or wave-like nature of light. We know now that this theoretical contradiction has been solved by QM \footnote{In hegelian terms, the wave packet is an example of ``Aufhebung'': contraries interpenetrate each other and overcome (``aufheben'') the contradiction} which takes into account what experiments teach us: corpuscular and wave-like aspects coexist and unite in the quantum behaviour of particles. A fact (a ``{\it{ funny logic}}''), that, seemingly, Nancy Cartwright, true to a classical aristotelian logic, is unable to accept.

\section{Conclusion}

In addition to questioning critically Nancy Cartwright's position on the measurement problem, on the wave function, and on the fundamental role she advocates for transition rates, the results mentioned above suggest rather convincingly that ``laws of physics'' (for example the Schroedinger equation) do tell truths about the world... Are such truths absolute ones? Given definite experimental conditions and technological stages of development, the answer cannot be but positive. For example the appearance of superconductivity below such and such temperature at ordinary pressure in such and such materials, or the appearance of quantum Hall effects, etc., not to speak of QM itself. So those absolute truths are also relative ones. Technological improvements, better accuracy of measurements,  different experimental parameters,  might  and most probably will enrich them by discoveries of new phenomena,  of new behaviours of matter, so that our representation of the world will evolve, without falsifying the absolute/relative truths mentioned above.  On the other hand, some truths have so to speak a higher degree of truth, inasmuch as they are true whatever happens  in the future: think of Noether's theorem, for example: the invariance of a system under a symmetry ensures the conservation of a corresponding entity in this system. 

I have argued that Nancy Cartwright does not identify correctly either the contradiction within QM, or  the (related)  dual nature of quantum evolutions (Schroedinger equation and reduction of the wave packet). She tries to argue it away by invoking an artefact of mathematics. This dual nature has its origin in the duality of our world where quantum objects and classical ones coexist and interact. The recent experiments I have described  confirm the Bohr Copenhagen interpretation of QM. There remains to establish more generally under what conditions an object behaves classically or according to quantum mechanics.

\vspace{.5cm}
{\bf{Acknowledgements.}} I wish to acknowledge useful and friendly exchange of e-mails with Franck Lalo\"e about the topics of this paper, and useful help from Jean-Michel Galano. I am grateful to Johanna Lederer for a careful correction of the manuscript.

\end{document}